\documentstyle[aps,epsf,twocolumn]{revtex}
\def\la{\langle} 
\def\ra{\rangle} 
\def\be{\begin{eqnarray}} 
\def\ee{\end{eqnarray}}

\newcommand{\eq}{\begin{equation}} \newcommand{\eqx}{\end{equation}}

\newcommand{\eqn}{\begin{eqnarray}} \newcommand{\eqnx}{\end{eqnarray}}

\begin{document}
\draft

\title{\bf Chiral Disorder and QCD Phase Transitions}

\author{ {\bf Romuald A. Janik}$^1$, {\bf Maciej A.  Nowak}$^{1,2}$ ,
{\bf G\'{a}bor Papp}$^{3}$ and {\bf Ismail Zahed}$^4$}

\address{$^1$ Department of Physics, Jagellonian University, 30-059
Krakow, Poland.
\\ $^2$ GSI, Planckstr. 1, D-64291 Darmstadt, Germany 
\\ $^3$ITP, Univ. Heidelberg, Philosophenweg 19, D-69120 Heidelberg, 
	Germany \& \\ Institute for
Theoretical Physics, E\"{o}tv\"{o}s University, Budapest, Hungary\\
$^4$Department of Physics and Astronomy, SUNY, Stony Brook, 
New York 11794, USA.}
\date{\today} \maketitle

\begin{abstract}
If QCD is to undergo a second order phase transition,
the light quark return probability is universal for large times at
the critical point. We show that this behavior is distinct from the one 
expected at the mobility edge of a metal-insulator transition or a percolation 
transition in d$\leq 4$. Our results are accessible to current lattice QCD 
simulations. 
\end{abstract}

\section{Introduction}

Light quarks trapped in a finite Euclidean volume $V$ behave like 
electrons in small metallic grains~\cite{USPRL}. The spontaneous 
breaking of chiral symmetry amounts to diffusing quarks in proper 
time with a vacuum diffusion constant 
$D\approx 0.22$ fm~\cite{USPRL}. For times larger than the ergodic
time $\tau_{\rm erg}=\sqrt{V}/D$ or small virtualities, 
the quarks undergo ergodic motion, while for times shorter than the elastic 
time $\tau_d=1/2m_Q\approx 0.33$ fm~\cite{USPRL} or large virtualities their 
motion is ballistic. Here $m_Q$ is the constituent quark mass in the vacuum.
Diffusion sets in at intermediate times and away from the quantum regime with
the Heisenberg time $t_H=1/\Delta=\Sigma V/\pi$~\cite{SMILGA}, and
$\Sigma$ the quark condensate.
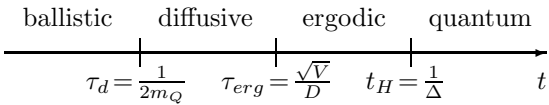
\begin{figure}[h]
\setlength{\unitlength}{1.2mm}
\begin{picture}(60,10)
\put(0,4){\vector(1,0){60}}
\put(15,2.5){\line(0,1){3}}
\put(30,2.5){\line(0,1){3}}
\put(45,2.5){\line(0,1){3}}
\put(9,0){\mbox{$\tau_d\!=\!\frac1{2m_Q}$}}
\put(24,0){\mbox{$\tau_{erg}\!=\!\frac{\sqrt{V}}D$}}
\put(40,0){\mbox{$t_H\!=\!\frac1{\Delta}$}}
\put(59,0){\mbox{$t$}}
\put(2,7){ballistic}
\put(17,7){diffusive}
\put(33,7){ergodic}
\put(47,7){quantum}
\end{picture}
\end{figure}

The concept of the quark return probability in the QCD vacuum as 
a chirally disordered medium~\cite{USPRL}, borrows on the concept
of the electron return probability in disordered metallic systems
as first introduced by Anderson~\cite{ANDERSON} in the context of
localization. For long times and in the presence of sufficient 
disorder, Anderson noted that the electron return probability 
is finite. A similar observation can be made for the quark return
probability in the QCD vacuum~\cite{USPRL}.

In this paper we would like to show that at the critical point of a second
order phase transition, a quantitative change in the character of the disorder
takes place, with consequences on the long time behavior of the
quark return probability. In section 2 and 3 we analyze the quark return 
probability in the vacuum and at the critical point. In section 4 and 5, we 
contrast our results to the expected ones from a metal-insulator and a 
classical percolation transition. In section 6, we use semi-classical 
arguments to show what this means for the 
light quark spectrum. Our conclusions are in section 7.

\section{Quark Return Probability}

The eigenvalue equation of the massless Dirac operator for
quarks in the fundamental representation and in the fixed gluon field $A$
\be
i\nabla \!\!\!\!/[A] \,q_k =\lambda_k [A] \, q_k
\label{01}
\ee
allows us to extend the theory into  4+1
dimensions  with  proper time $t$,
and define the probability $p(t)$ for a light quark to start at $x(0)$
in $V$ and return back to the same position $x(t)$ after a duration $t$, as
\be
p(t)= \frac {V^2}N
\Big\la |\la x(0)|e^{i(i\nabla \!\!\!\!/[A] +im)|t|}|x(0)\ra|^2\Big\ra_A \,.
\label{1}
\ee
The operator $i\nabla \!\!\!\!/[A]$ in~(\ref{01}) acts as a four-dimensional 
Hamiltonian for the evolution in proper time $t$. Indeed, the
expectation value in (\ref{1}) is  that of the 
proper time evolution operator, with the background gluon field $A$
acting as a t-independent (static) random potential.
The averaging in (\ref{1}) is over all gluon 
configurations using the unquenched QCD measure, with
light quark flavors of current mass $m$ (flavor symmetric case). The
normalization in (\ref{1}) is per state, where $N$ is the mean number
of quark states in the four-volume $V$.

In terms of the quark-eigenfunctions
(\ref{01}) in $V$, the return probability reads
\be
p(t) =\frac {V^2}{N}&& e^{-2m|t|} \sum_{j,k} \nonumber\\
&&\times\Big\la e^{i|t| (\lambda_j-\lambda_k)[A]} |q_j (x)|^2 |q_k (x)|^2\Big\ra_A
\label{02}
\ee
where the exponent $e^{-2m|t|}$ is solely due to the current quark mass
in (\ref{1}). We note that (\ref{02}) is gauge-invariant and 
amenable to lattice QCD simulation. This form is best suited for numerical 
estimates.

For analytical considerations, it is best to rewrite (\ref{1}) in terms
of the standard Euclidean propagators for the quark field,
\be
p(t) = \frac {V^2}N \lim_{y\to x}{}&&
\int \frac {d\lambda_1d\lambda_2}{(2\pi)^2} 
\,e^{-i(\lambda_1-\lambda_2) |t|}\nonumber\\&&
\times\Big\la {\rm Tr}\left( S(x,y;z_1) S^{\dagger} (x,y; z_2)\right)\Big\ra_A
\label{des2}
\ee
with $z_{1,2}=m-i\lambda_{1,2}$, and
\be
S(x,y; z) = \la x| \frac 1{i\nabla \!\!\!\!/[A] + iz} |y\ra \,.
\label{des3}
\ee
Setting $\lambda_{1,2}=\Lambda\pm \lambda/2$ and neglecting the effects
of $\Lambda$ in the averaging in (\ref{des2}) (this is certainly true
near the zero virtuality point), we find that in the flavor symmetric
limit, the correlation function in (\ref{des2}) relates to the pion 
correlation function for a proper analytical continuation of the 
 current quark mass~\cite{USPRL}. Specifically,
\be
p(t) = \frac {EV^2}{2\pi N} \lim_{y\to x}{}
\int \frac {d\lambda}{2\pi} 
\,e^{-i\lambda |t|} {\bf C}_{\pi} (x,y; z)
\label{des55}
\ee
where
\be
\label{des5}
\,{\bf 1}^{ab}\, {\bf C}_{\pi} (x,y;z) &=& \\ 
&&\hspace*{-6mm} \Big\la {\rm Tr}\left(
S(x, y;z) i\gamma_5\tau^a S(y,x;z) i\gamma_5\tau^b\right)\Big\ra_A
\nonumber
\ee
with $z=m-i\lambda/2$ and $E=\int d\Lambda$.
For $z=m$, pion-pole dominance (long paths) yields
\be
{\bf C}_{\pi} (x,y;m) \approx \frac 1V \sum_Q e^{iQ\cdot (x-y)} 
\frac {\Sigma^2}{F^2} \frac 1{Q^2+m_{\pi}^2}
\label{des06}
\ee
with $Q_\mu =n_{\mu}2\pi/L$ in $V=L^4$ and $\Sigma=|\la \overline{q} q\ra |$.

Using the GOR relation $F^2m_{\pi}^2=m\Sigma$, 
and the analytical continuation $m\rightarrow m-i\lambda/2$, we 
find~\cite{USPRL}
\be
{\bf C}_{\pi} (x,y;z) \approx \frac 1V \sum_Q e^{iQ\cdot (x-y)} 
\frac {2\Sigma}{-i\lambda + 2m + DQ^2}
\label{des6}
\ee
with the diffusion constant $D=2F^2/\Sigma$. 
Inserting (\ref{des6}) into (\ref{des55}), and noting that
$E/\Delta=N$ and $\rho=1/\Delta V$, where $\Delta$ is the mean quantum
level spacing
with $\Sigma=\pi \rho$ (Banks-Casher relation), we conclude after
a contour integration that
\be
p(t) = e^{-2m|t|}\sum_Q e^{-DQ^2 |t|} \,.
\label{des7}
\ee
The validity of (\ref{des7}) is for $\tau_d<t<t_H$. For $m\,t_H\sim mV\ll 1$,
(\ref{des7}) is dominated by the constant pionic mode, giving $p(t)\sim 1$. 
What happens to the present arguments if QCD is to undergo a phase transition?

\section{Scaling and Universality}

During a second order transition, the quark condensate and the pion 
parameters undergo structural changes which are governed by scaling 
arguments~\cite{KOCIC}. For $z=m$ (\ref{des06}) takes the general 
form
\be
{\bf C}_{\pi} (x,y;m) \approx \frac 1V \sum_Q e^{iQ\cdot (x-y)} 
\frac {Z_{\pi}}{Q^2+Z_{\pi}/\chi_{\pi}}
\label{sca0}
\ee
where $Z_{\pi}$ is the pion wavefunction renormalization, and 
$\chi_{\pi}$ the pion susceptibility. At the critical point $T_C$
(generic of a second order transition), $V=L^3/T_C$. The representation
(\ref{sca0}) holds for long paths, or Euclidean momenta of order
$\Lambda\leq (2\pi T_C)$. For larger Euclidean momenta or short paths, 
(\ref{sca0}) is saturated by quarks in the dimensionally
reduced theory~\cite{HANS}.

At the critical point, we have~\cite{KOCIC}
\be
Z_{\pi} =&&\Sigma^{\nu\eta/\beta}\nonumber\\
\chi_{\pi}=&&\Sigma^{1-\delta}\nonumber\\
\Sigma=&&m^{1/\delta}
\label{sca1}
\ee
where $\beta,\nu,\delta,\eta$ are critical exponents. The prefactors
in (\ref{sca1})  are dimensionful constants that relate to the equation of 
state (the last equation). For convenience they were set to 1. Hence,
\be
{\bf C}_{\pi} (x,y;m) \approx \frac {1}{V}
\sum_Q e^{iQ\cdot (x-y)}  \frac {m^A}{Q^2+ m^B}
\label{sca2}
\ee
where $A=B-1+1/\delta$ and $B=4/\delta(2+\eta)$ after using scaling 
relations. 
For a mean-field transition there is no wave-function renormalization
and $\eta=0$. Therefore: $\delta=3$, $A=0$ and $B=2/3$. After the analytical 
continuation $m\rightarrow  m-i\lambda/2$, we obtain
\be
{\bf C}_{\pi} (x,y;z) \approx \frac {1}{V }
\sum_Q e^{iQ\cdot (x-y)}  \frac {(m-i\lambda/2)^A}{Q^2+  
(m-i\lambda/2)^B} \,.
\label{sca4}
\ee
As a result,  the quark return probability becomes
\be
p(t)= {\bf C} \sum_Q
\int \frac{d\lambda}{(2\pi)^2} 
{e^{-i\lambda |t|}}\frac {(2m-i\lambda)^A}{2^B{Q}^2+ 
(2m-i\lambda)^B}
\label{sca5}
\ee
where ${\bf C}=VE/N2^{A-B}$.

In contrast to the vacuum case (\ref{des6}), the result 
(\ref{sca4}) at the critical point displays new singularities.
For fixed $Q^2$, the pole in the $\lambda$-plane in (\ref{des6}) 
characteristic of the vacuum phase is now changed to a branch point 
at $\lambda=-2im$, and a set of poles ($Q\neq 0$)
\be
\lambda_n=-2im +2ie^{i(2n+1)\pi/B} 
{|{ Q}^2|}^{1/B}
\label{sca6}
\ee
with $n=0, 1, ..., n_{\rm max}$, and $n_{\rm max}$ is the 
number of unit roots to $1+z^B=0$. The contribution of the cut is
\be
p_c(t)=&&-\frac{VE}{N\pi^22^{1/\delta+1}}e^{-2m|t|}\nonumber\\
&&\times \sum_Q \int_0^{\infty} dx\, 
e^{-x|t|}\,\, 
{\rm Im}\left( \frac{{x^A}e^{i\pi A}}{2^B Q^2+x^Be^{i\pi B}}\right)
\label{sca07}
\ee
while the contribution of the poles is
\be
p_p (t)= - \frac {VE}{\pi\,BN}\sum_{Q,n}'
e^{-i\lambda_n |t|} \left(m-i\lambda_n/2\right)^{1/\delta} \,.
\label{sca08}
\ee
The primed sum in (\ref{sca08}) retains 
only those poles in the lower half of the $\lambda$-plane.
The result is still real as the poles occur in $(\lambda,-\lambda^*)$
pairs.

For large times, the dominant contribution to $p(t)$ results 
from the $Q=0$ part (zero mode) of (\ref{sca07}). Setting the Heisenberg
time to be $t_H=1/\Delta_*\sim V^{\delta/(1+\delta)}$~\cite{PRLUS1}, and
the mean number
of levels to be $N=E/\Delta_* \gg 1$, the quark return probability simplifies
\be
p(t) \approx \frac {1}{2\pi^2} {\rm sin}({\pi}/{\delta})
\Gamma (1/{\delta} )\,\,{e^{-\alpha_*|t/t_H|}} 
\left( \frac {t_H}{2t}\right)^{1/\delta} \,.
\label{sca7}
\ee
For $\alpha_*=2mV^{\delta/(\delta+1)}\ll 1$ (ergodic regime), the 
return probability for large times $t$ is universal at the critical point 
with $p(t)\sim (t_H/t)^{1/\delta}$. This behavior
is to be contrasted with the vacuum result $p(t)\sim 1$ in 
the ergodic regime  and $(\tau_{\rm erg}/t)^2$ in the diffusive 
regime~\cite{USPRL}.

We note that (\ref{sca4}) obeys an `anomalous' diffusion equation in d=4
at $m=0$ (with the cutoff $\Lambda$)
\cite{NOTE4},
\be
\left[- D(\lambda ) \nabla^2 + (-i\lambda )\right]\,
C_{\pi} (x, y; -i\lambda )=0
\label{diff}
\ee
for $x\neq y$. At the critical point the diffusion constant
is $\lambda$ dependent and complex $D(\lambda )=2^{B}(-i\lambda)^{1-B}$.
For $\lambda \sim m\sim 0$ (zero virtuality)
the diffusion constant vanishes as 
$|D(m)|\sim m^{1-B}$. In this case, the Heisenberg time is 
$t_H=1/\Delta_*\sim V^{\delta/(\delta +1)}$,
and the ergodic time is $\tau_{\rm erg}=1/E_c\sim \sqrt{V}/m^{1-B}$,
where $E_c$ is the Thouless energy~\cite{ASYM}.
Since the constituent quark mass (half the sigma mass) becomes
comparable to the pion mass at the critical point, the elastic
time is $\tau_d\sim 1/m^{B/2}$ (see~(\ref{sca2})). 
To assess the hierarchy of scales near the critical point, we enhance
artificially the contribution from the ergodic regime by setting
$m/\Delta_*\sim 1$ in power counting, in generalization of a previous
argument~\cite{GASSERL}. For a mean-field transition: $\tau_d\sim V^{1/4}$ 
and $\tau_{\rm erg}\sim t_H\sim V^{3/4}$. Hence the Ohmic conductance
$\sigma_L=t_H/\tau_{\rm erg}\sim 1$ a situation reminiscent of the
metal-insulator transition, except that in the present case the density
of states vanishes as $\lambda^{1/\delta}$. The time scales are still 
ordered in the thermodynamical limit, with the diffusive regime 
stretching all the way to the quantum regime.

We note  that the same scaling arguments
we have used at $T=T_c$ also indicates that near but above the critical
temperature $T>T_c$, the pion mass vanishes in the symmetric phase in
general with $m_{\pi}^2\sim \Sigma^{\eta\nu/\beta}$ as $\Sigma\sim 
m$~\cite{KOCIC}, owing to wave-function renormalization~\cite{NOTE1}. 
Some of our previous arguments may apply provided that due care is paid to 
the pion dispersion relation with $Q^2\rightarrow Q^{2-\eta}$~\cite{KOCIC}.
Similar behavior has been noted at the mobility edge of a metal-insulator 
transition in metallic grains~\cite{KRAVSTOV}.

\section{Metal-Insulator Transition}

How does the present chiral phase transition compare
to the metal-insulator (MI) transition in QCD in vacuum we discussed
recently in d=4~\cite{USPRL}? For comparison, we note that in the latter and
for long times the quark return probability scales as~\cite{USPRL,CHALKER}
\be
p(t)\approx \frac{{e^{-2m |t|}}}{{|t|^{1-{\eta/4}}}}
\label{frac1}
\ee
with a multifractal exponent~\cite{USPRL}
\be
\frac {\eta}4 = 2\chi
=4\frac {1-0.577+{\rm ln 4}}{\beta (2\pi)^4} \frac 1{\sigma_*} \,.
\label{frac2}
\ee
In QCD $\beta=2$ and the critical conductance $\sigma_*$ is equal to
the microscopic conductance $\sigma_l =2F^2l^2/\pi$ for one mean-free 
path $l$, that is~\cite{USPRL,NOTE2}
\be
\sigma_l = \frac 8{\pi} \frac {F^4}{\Sigma m_Q}\approx 0.041
\label{frac3}
\ee
where $F\approx 93$ MeV is the pion decay constant, $\Sigma\approx (250)$ 
(MeV)$^3$ is the vacuum quark condensate and $m_Q\approx 300$ MeV a typical
constituent quark mass. Hence $\eta/4\approx 0.057$, 
which yields a larger critical exponent than in (\ref{sca7}). We note that 
our prediction for the critical exponent in QCD is in 
qualitative agreement with the numerical estimates $\eta/4\approx 0.08-0.16$ 
obtained subsequently using the instanton liquid model~\cite{OSB}. Remarkably,
the smallness of $\eta/4$ makes the result (\ref{frac1}) close to a 
2-dimensional diffusion process.

In the MI transition, (\ref{frac3}) follows from the ballistic region and is 
related to a careful consideration of the null spectral sum 
rule~\cite{USPRL,KRAVSTOV,BALLISTIC}. At the critical point 
this is necessary because of the multifractal character of the 
corresponding wavefunctions at the mobility edge~\cite{CHALKER}.
In the metallic regime, the quark wavefunctions
are spread through the metal, while in the insulator regime they are 
localized. At the edge, the wavefunctions are `filamentary' with self-similar
structure~\cite{AOKI}. In a second order transition we expect the quark 
wavefunctions to be persistently `metallic'.

\section{Percolation Transition}

Multifractal wavefunctions have been deemed as `quantum' percolation by 
Aharony~\cite{AHARONY}. 
Could a `classical' percolation transition take place at finite
temperature in QCD? Recently, some arguments have been put forward by
Satz~\cite{SATZ}. In the present framework this can be
addressed by noting that (\ref{sca0}) valid at $T_C$ by standard 
scaling arguments~\cite{KOCIC} does not incorporate the fact that 
for temperatures $0<T<T_C$ the Goldstone modes disperse asymmetrically in 
matter~\cite{PISARSKI,TOUBLAN}. Indeed, in these range of temperatures 
and for space-like momenta we have instead of (\ref{des6}),
\be
{\bf C}_{\pi} (x,y,z)\approx &&\frac 1V\sum_Q
e^{iQ\cdot(x-y)}\nonumber\\&&\times\frac{2\Sigma_T}
{D_4 Q_4^2 + D_S \vec{Q}^2 +2m -i\lambda}
\label{as1}
\ee
with $\Sigma_T=|\la \overline{q} q\ra_T|$ the temperature dependent
quark condensate. Here,
the `temporal' $D_4$ and `spatial' $D_S$ diffusion constants
are related to the `temporal' $F_4$ and `spatial' $F_S$ weak decay
constants of the pion~\cite{PISARSKI}. In the space-like regime they 
are both real, and we have generically
\be
&&D_4=2{F^2_4}/{\Sigma_T}\,,\nonumber\\
&&D_S=2{F_4F_S}/{\Sigma_T}\,.
\label{as2}
\ee
If we were to denote by $m_4^2$ the squared pion `time-like' mass,
and by $m_S^2$ the squared pion `space-like' mass,
then $D_4=2m/m_4^2$ and $D_S=2m/m_S^2$ where $m_S\geq m_4$ by
causality~\cite{PISARSKI}. Hence $D_4\geq D_S$. 

In terms of (\ref{as1}) the quark return probability in the range
$0<T<T_C$ reads
\be
p(t) = e^{-2m |t|}\sum_Q e^{-D_4Q_4^2 |t|-D_S\vec{Q}^2 |t|}
\label{as03}
\ee
where we have used the facts that $\Sigma_T=\pi\rho_T$, $\Delta_T=1/\rho_TV$
and $N=E/\Delta_T$. At finite temperature both the density of states $\rho_T$
and the level spacing $\Delta_T$ change in a 4-volume $V=L^3/T$. The momenta
in (\ref{as03}) are $Q_4=n_42\pi T$ and $\vec{Q}=\vec{n}2\pi/L$. 
(\ref{as03})
implies the existence of two Thouless energies $E_4=D_4T^2$ and $E_S=D_S/L^2$,
hence two ergodic times $\tau_{4,{\rm erg}}=1/E_4$ and $\tau_{S,{\rm
erg}}=1/E_S$, with $\tau_{4,{\rm erg}}<\tau_{S, {\rm erg}}$. The
corresponding Ohmic conductances 
are $\sigma_4=E_4/\Delta_T$ and $\sigma_S=E_S/\Delta_T$. These scales allow
a simple organization of the disorder in an asymmetric (finite temperature)
Euclidean box. In particular, (\ref{as03}) becomes universal for $mV\ll 1$
and times $t\gg \tau_{S, {\rm erg}} > \tau_{4, {\rm erg}}$. Most of the vacuum
arguments presented in~\cite{USPRL} can be extended to the present finite
temperature phase.

If we were to denote the `temporal' $\varrho_4$ and `spatial' $\varrho_S$
quark conductivities, then by the Kubo formulae: $\varrho=D\rho_T$, we have
$\varrho_4\geq \varrho_S$. 
For temperatures $0<T<T_C$ the quark conductivity in the `spatial' directions
is weaker than the quark conductivity in the `temporal' direction. This
is easily understood by the fact that `spatially' the system `screens' all 
charges including the singlet ones. Indeed, the pion screening length at high
temperature asymptotes $m_S^2\sim2\pi T$ by standard
arguments~\cite{HANS}, so that the 
spatial conductivity $D_S\sim 2m/(2\pi T)^2$ is parametrically small in the 
high temperature phase. 

The asymmetry in the conduction properties may cause an `asymmetric'
percolation from d=4 to d=1 as $D_S$ becomes much smaller
than $D_4$, a situation reminiscent of finite density~\cite{PREPARA}. 
{}From (\ref{as03}), it follows that
\be
p(t,T)\approx\frac{e^{-2m|t|}}{\sqrt{4\pi E_4 |t|}}
\label{as003}
\ee
in the limit $D_4\gg D_S$. 
We note with Aharony~\cite{AHARONY} that in a 
`classical' percolation transition in either d=3 or d=4,
the quark return probability on an infinite critical
cluster is expected to scale as
\be
p(t)\approx \frac{e^{-2m|t|}}{{|t|^{{\tilde{D}}/{(2+\theta)}}}}
\label{frac1a}
\ee
where $\tilde{D}=2.5,3.2$ is the fractal dimension and $\theta=1.77, 2.70$ in d=3
and d=4 respectively. Hence, $\tilde{D}/(2+\theta)=0.66, 0.68$ for d=3,4
respectively, which are similar to 0.5 in (\ref{as003}). Both exponents  are 
markedly different from the MI behavior~(\ref{frac1}).

In a linear sigma model the diffusion
constants $D_4$ and $D_S$ can be estimated in a weak-coupling (1/n) expansion
\cite{PISARSKI}. The results in our case and to 1-loop are
\be
&&D_4\sim D\left(1-\frac{T^2}{24F^2}+\frac {6\pi^2}{45} \frac 
{T^4}{F^2m_{\sigma}^2}\right)\,,\nonumber\\
&&D_S\sim D\left(1-\frac{T^2}{24F^2}-\frac {2\pi^2}{45} \frac 
{T^4}{F^2m_{\sigma}^2}\right)
\label{as3}
\ee
where $m_{\sigma}\sim 500$ MeV is a typical sigma mass. 
To the same-order we have $\Sigma_T\sim\,\Sigma \,(1-T^2/8F^2)$.
Similar calculations can be performed in the context of the non-linear 
sigma model as well using instead 2-loops. A fine tuning of the running
logarithms in this case~\cite{TOUBLAN} should bring both estimations 
into agreement. From (\ref{as3}) it follows that an order of magnitude drop
in the relative conductivities $\varrho_S/\varrho_4\sim 0.1$ or equivalently
\be
\frac {D_S}{D_4}\sim 1-\frac{8\pi^2}{45} \frac{T^4}{F^2m_{\sigma}^2}\sim 0.1
\label{ratio}
\ee
takes place for $T\sim 180$ MeV, which is were we expect the 
percolation transition to possibly set in. At this temperature only 
$50\%$ of the chiral condensate or the density of quark states at zero 
virtuality has depleted by the 1-loop argument. It would be interesting to 
investigate the effects of strangeness on the present arguments.

\section{Semi-Classical Approximation}

What does the result (\ref{sca7}) mean for the light quark spectrum in QCD?
A qualitative answer to this question can be obtained using semi-classical
arguments~\cite{IMRY,MONTAMBAUX}. Since in our case the density of states
$\rho (\lambda)\sim \lambda^{1/\delta}$ changes in the energy band of interest,
the standard semi-classical arguments have to be modified. Indeed, a rerun of 
the standard arguments~\cite{IMRY,MONTAMBAUX} yields for the quark return 
probability 
\be
p(t)=\frac {V^2}{N}\Bigg\la \sum_j |A_j (x)|^2 \,\delta(t-T_j)\Bigg\ra_A
\label{sc1}
\ee
and the spectral form-factor 
\be
K(t)=\frac 1{2\pi^2}\Bigg\la\sum_j \left|\int\!dx \, A_j (x)\right|^2 \, 
\delta(t-T_j)\Bigg\ra_A
\label{sc2}
\ee
where the sum is over closed classical paths of reduced action $S_j 
(\Lambda)$, virtuality $\Lambda$, and period $T_j=dS_j/d\Lambda$.
$A_j (x)$ is the Gaussian contribution around the
classical paths to the propagator $S(x,x;m-i(\Lambda\pm \lambda/2))$. 
The averaging in (\ref{sc2}) is over the gauge-configurations 
(static or t-independent disorder).

The contributions to  (\ref{sc1}) result from closed
classical paths labeled by $j$, starting at $x$ and returning back to $x$
with probability $P_j=|A_j(x)|^2$. The contributions to~(\ref{sc2}) differ
from those of~(\ref{sc1}) by the overlap integration over the `4-volume' 
of a given classical path~\cite{IMRY,MONTAMBAUX}.
For a virtuality $\lambda_j\sim 1/T_j$ about the mean $\Lambda\sim 0$,
the `4-volume' is of order $T_j/\rho (1/T_j)$, as $\rho (\lambda)$ counts
the number of states per `4-volume' per virtuality.
Inserting this result into (\ref{sc2}) and using (\ref{sc1}), we conclude that
\be
K(t)\approx {\bf K}\, \frac{\Delta^2_*|t|^{1+1/\delta}}{(2\pi )^2\beta}
 \, p(t)
\label{sc3}
\ee
The dimensionless coefficient ${\bf K}$ is not fixed by these estimates. 
For QCD $\beta=2$~\cite{USPRL}. 
(\ref{sc3}) in combination with (\ref{sca7}) yields $K(t)\sim |t|$
for large times ($m=0$). This result at $T=T_c$ is reminiscent of the result
at $T=0$ and suggests once more that it is universal. Hence, the corresponding
spectral rigidity is still logarithmic but with a coefficient that is not 
fixed by the present semi-classical arguments. It can be fixed using our
recent arguments on critical scaling in 0-dimension~\cite{PRLUS1,PRLUS2}.

\section{Conclusions}

We have shown that the long time behavior of the quark return 
probability in a QCD phase transition carries a quantitative information 
on the character of a phase change, that could be used to discriminate 
between a second order, a metal-insulator or a percolation transition.
For times comparable to the Heisenberg time $t_H\sim 1/\Delta_*$
and in the narrowing ergodic regime $m\ll\Delta_*$, we have found that the 
quark return probability at the critical point scales as $|t|^{-1/\delta}$ 
for a second-order transition. In contrast, in a metal-insulator
transition it is of order $|t|^{-0.943}$ while for an asymmetric percolation 
transition it is of order $|t|^{-1/2}$ from d=4 to d=1. Our results can
be tested by numerically assessing (\ref{02}) using current lattice QCD 
algorithms.

\vskip 0.5cm
{\bf Acknowledgments}
\\

IZ would like to thank Igor Aleiner for a discussion.
This work was supported in part by the US DOE grant DE-FG-88ER40388, by the 
Polish Government Project (KBN) grants 2P03B04412 and 2P03B00814 and by the 
Hungarian grants FKFP-0126/1997 and OTKA-F026622.
RJ was supported by the Foundation for Polish Science.


\begin{thebibliography}{99}
\bibitem{ASYM}
Due to the asymmetry of the 4-volume at finite
temperature there are in fact two distinct Thouless energies,
corresponding to diffusion through the sample in spatial and temporal
directions. The spatial size $L$ and the (critical) temporal size $1/T_C$ 
bring about various combinations of ergodic/diffusive behavior, in
particular a possible 3-dimensional diffusive regime with temporal
`ergodicity'. We will discuss some of these issues in section 5.
\bibitem{NOTE2}
It is amusing to note that the RGE analysis in $2+\epsilon$ 
dimensions~\cite{LEE} yields $\sigma_*=1/2\pi^2$ which is rather 
close to the present result.
\end{thebibliography}
\end{document}